\documentclass[twoside,11pt]{article}
\usepackage{jmlr2e}

\ShortHeadings{Causal Discovery Toolbox}{Kalainathan et al.}

\firstpageno{1}

\usepackage[utf8]{inputenc}
\usepackage{todonotes}
\usepackage{hyperref}       
\usepackage{url}            
\usepackage{booktabs}       
\usepackage{amsfonts}       
\usepackage{graphicx}
\usepackage{amsmath}
\usepackage{stmaryrd}
\usepackage{amsfonts}
\usepackage{subcaption}
\usepackage{pbox}
\usepackage{array}
\usepackage{hyperref}
\usepackage{geometry}
\usepackage[ruled,vlined]{algorithm2e}
\usepackage{natbib}
\usepackage{graphicx}
\usepackage{smartdiagram}
\graphicspath{{./figures/}}
\usepackage{tikz}
\usetikzlibrary{arrows.meta,
                chains,
                positioning,
                shapes.geometric
                }

\newcommand*{\indep}{%
  \mathbin{%
    \mathpalette{\@indep}{}%
  }%
}

\title{Causal Discovery Toolbox:
\\ \Large Uncover causal relationships in Python}
\author{\name Diviyan Kalainathan \email diviyan.kalainathan@inria.fr\\
\addr TAU, LRI, INRIA, Université Paris-Sud\\
660 Rue Noetzlin, 91190 Gif-Sur-Yvette, France\\
\AND
\name Olivier Goudet \email olivier.goudet@univ-angers.fr\\
\addr LERIA, Université d’Angers,\\
2 boulevard Lavoisier, 49045 Angers, France\\
}
\date{}
\editor{}
\def\cdt{{\sc Cdt}}
\begin{document}
\maketitle
\begin{abstract}
  This paper presents a new open source Python framework for causal discovery from observational data and domain background knowledge, aimed at causal graph and causal mechanism modeling. The \textbf{\cdt} package implements the end-to-end approach, recovering the direct dependencies (the skeleton of the causal graph) and the causal relationships between variables. It includes algorithms from the '\textbf{\sc Bnlearn}' \citep{scutari2018package} and '\textbf{\sc Pcalg}' \citep{kalisch2018package} packages, together with algorithms for pairwise causal discovery such as ANM \citep{hoyer2009nonlinear}. \textbf{\cdt} is available under the MIT License at \url{https://github.com/Diviyan-Kalainathan/CausalDiscoveryToolbox}.

\end{abstract}

\section{Introduction}
Causal modeling is key to understand physical or artificial phenomenons and make recommendations. Most softwares for causal discovery have been developed in the R programming language \citep{kalisch2018package, scutari2018package}, and a few causal discovery algorithms are available in Python (RCC \citep{lopez2015towards}, CGNN \citep{goudet2017learning} and SAM \citep{kalainathan2018sam}), which supports many current machine learning frameworks such as PyTorch \citep{paszke2017automatic}. 

The proposed \cdt\ package is concerned with observational causal discovery, aimed at learning both the causal graph and the associated causal mechanisms from samples of the joint probability distribution of the data. \cdt\ is supported by PyTorch.

Formally, the \textbf{Causal Discovery Toolbox} (\cdt) is a open-source Python package including many state-of-the-art causal modeling algorithms (most of which are imported from R), that supports GPU hardware acceleration and automatic hardware detection. 

\begin{figure}[h]
\centering
\includegraphics[width=.8\textwidth]{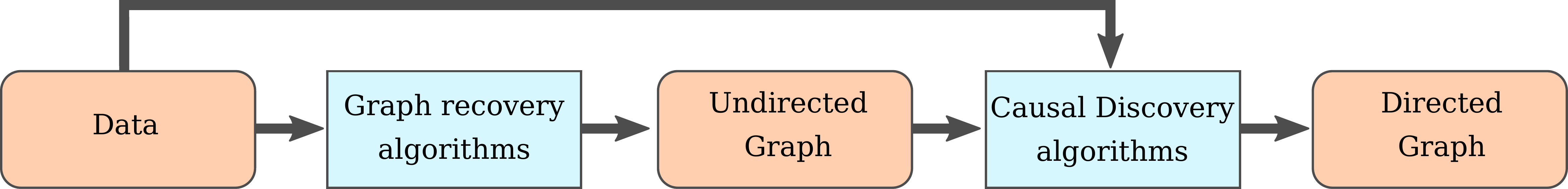}

\caption{The \cdt\ causal modeling package: General pipeline\label{fig:pipeline}}
\end{figure}

Compared to other causal discovery packages, \cdt\ unifies pairwise and score-based multi-variate approaches within a single package, implementing an end-to-end, step-by-step pipeline approach (Fig. \ref{fig:pipeline}).

\cdt\ also provides an intuitive approach for including R based algorithms, facilitating the task of extending the toolkit with additional R packages. The package revolves around the usage of \textit{networkx.Graph} classes, mainly for recovering (un)directed graphs from observational data.

\section{Algorithms}
  \cdt\ currently includes 17 algorithms for graph skeleton identification: 7 methods based on independence tests, and 10 methods aimed at directly recovering the skeleton graph. It furter includes 19 algorithms aimed at causal directed graph prediction, including 10 graphical and 9 pairwise approaches.

\subsection{Recovering the graph skeleton}
\cdt\ includes two types of methods for recovering undirected dependence graphs from raw data: methods based on pairwise dependence statistics (also referred to as bivariate methods), and methods based on variable/feature selection.
\paragraph{Bivariate dependencies} support variable/feature selection and are used to determine the (undirected) edges in the causal graph. They rely on statistical tests, e.g. Pearson's correlation or mutual information scores \citep{vinh2010information}. 
Bivariate dependencies are used in a first phase to establish the causal graph skeleton. In a further phase, heuristics aimed at builing a (causal) DAG from the causal graph skeleton are used. In particular, indirect edges (that is, edge $A \rightarrow C$, when edges $A\rightarrow B$ and $B \rightarrow C$ have been established) are removed using e.g.,  Network Deconvolution \citep{feizi2013network}. These graph pruning heuristics can be parameterized to control the sparsity of the graph.

\paragraph{Multivariate methods} aims at recovering the full causal graph, that is, selecting parent, children and spouse nodes (parents of children) for all variables of the graph. This task has been thoroughly investigated through feature selection, graph heuristics \citep{friedman2008sparse}, and Markov blankets \citep{tsamardinos2003algorithms}. All these methods output a \url{networkx.Graph} object.

\subsection{Causal discovery}
The main focus of the \cdt\ package is causal discovery from observational data, ranging from the pairwise setting to the full graph modeling. 

\paragraph{The pairwise setting} considers a pair of variables and aims to determine the causal relationship between those variables. This setting implicitly assumes that both variables are already conditioned on other covariates, or readjusted with a propensity score \citep{rosenbaum1983central}, and that remaining latent covariates have little or no influence and can be considered as ``noise". The pairwise setting is also relevant to complete a partially directed graph resulting from other causal discovery methods. In the 2010s, the pairwise setting was investigated by \cite{hoyer2009nonlinear} among others, respectively proposing the Additive Noise Model (ANM). Later on, \citep{guyon2013cepc}  launched international challenges on Kaggle and Codalab on Cause-Effect pair (CEP) problems; CEP formulates bivariate causal identification as a machine learning task, where a classifier is trained from examples ($A_i,B_i,\ell_i$), where variable pair ($A_i,B_i$) is represented by samples of their joint distribution and label $\ell_i$ indicates the type of causal relationship among both variables (independent, $A_i \rightarrow B_i$, $B_i \rightarrow A_i$). The CEP challenges spurred the development of state-of-the-art pairwise causal identification methods such as Jarfo, RCC \citep{fonollosa2016conditional, lopez2015towards}.

\paragraph{The graph setting,} extensively studied in the literature, involves Bayesian or score-based approaches. Bayesian approches rely either on conditional independence tests named \textbf{constraint-based methods}, such as PC or FCI \citep{spirtes2000causation, strobl2017approximate}, or on \textbf{score-based methods}, finding the graph that maximizes a likelihood score through graph search heuristics, like GES or CAM \citep{chickering2002optimal, buhlmann2014cam}. Other approaches leverage the celebrated Generative Adversarial
Network setting \cite{GAN}, such as CGNN or SAM \citep{goudet2017learning, kalainathan2018sam}. Graph setting methods
output either a directed acyclic graph or a partially directed acyclic graph.

\section{Implementation and utilities}
\paragraph{R integration.}
\label{subsec:settr}
As said, currently 10 algorithms are coded in R, and 17 in Python. The \cdt\ package integrates all of them, using Wrapper functions in Python to enable the user to 
launch any R script and to control its arguments; this R script is executed in a temporary folder with a \textit{subprocess} to avoid the limitations of the Python GIL. The results are retrieved through files back into the main Python process. The whole procedure is modular and allows contributors to easily add new R functions to the package.

\paragraph{Hardware configuration settings.}
\label{subsec:setthw}
At the package import, tests are realized to pinpoint the configuration of the user: Availability of GPUs and R packages and number of CPUs on the host machine. All settings are stored in a single object \url{cdt.SETTINGS}. For some algorithms, GPUs accelerations are available through the use of the PyTorch library. 

\paragraph{Sustainability and deployment.} In order for the package to be easily extended, integrating and encouraging the community contributions, special care was paid to the quality of tests. Specifically, a Continuous Integration tool (CIT),  added to the git repository, allows one to sequentially execute tests. On new commits and pull request, the CIT automatically:
i) Test all functionalities of the new version on the package using \emph{pytest}\footnote{Holger Krekel et al., \url{https://github.com/pytest-dev/pytest/}} on toy datasets;
ii) Build docker images and push them to \url{hub.docker.com} ;
iii) Push the new version on \emph{pypi};
iv) Update the documentation website. 
This procedure also allows to test the proper functioning of the package with its dependencies.

\section{Conclusion and future developments}
The {Causal Discovery Toolbox} (\cdt) package (\cdt) allows Python users to apply many causal discovery or graph modeling algorithms on observational data. It is already used by \citep{goudet2017learning, kalainathan2018sam}. As the output graphs are \url{networkx.Graph} classes, these are easily exportable into various formats for visualization such as Graphviz and Gephi. 

The package promotes an end-to-end, step-by-step approach: the undirected graph (bivariate dependencies) is first identified, before applying causal discovery algorithms; the latter are constrained from the undirected graph, with significant computational gains. \par
Future extensions of the package include: i) developing GPU-compliant implementation of new  algorithms; ii) handling interventional data and time-series data (e.g. for neuroimaging and weather forecast); 
iii) evaluating the direct and total effect in a graph, given a cause variable and a target variable. 
Finally, we plan to develop facilities to test whether common assumptions (e.g. causal sufficiency assumption) hold and reduce the risk of applying methods out of their intended scope. 

\bibliography{bibliography}

\end{document}